             \documentclass{aastex631}
\usepackage{color}
\begin{document}

%--------------------------------------
%                                                   \input{"$HOME/science/LateX_newcommands.txt"}   %   <================= new LateX commands 
%---------------------------------------

\title{THE EXTERNAL HEATING OF DUST \\ IN A HOMOGENEOUS SPHERICAL SHELL}

\author[0000-0001-8033-1181]{Eli Dwek}
\affiliation{Emeritus, Observational Cosmology Lab, NASA Goddard Space Flight Center, Mail Code 665, Greenbelt, MD 20771, USA}
\email{elidwek.astro@gmail.gov}
\affiliation{Research Fellow,  Center for Astrophysics | Harvard \& Smithsonian, 60 Garden Street, Cambridge, MA 02138, USA}

\author[0000-0001-8403-8548]{Richard G. Arendt}
\affiliation{Center for Space Sciences and Technology, University of Maryland, Baltimore County, Baltimore, MD 21250, USA}
\affiliation{Code 665, NASA/GSFC, 8800 Greenbelt Road, Greenbelt, MD 20771, USA}
\affiliation{Center for Research and Exploration in Space Science and Technology, NASA/GSFC, Greenbelt, MD 20771, USA}

%\author[0000-0002-9820-679X]{Arkaprabha Sarangi}
%\affiliation{DARK, Niels Bohr Institute, University of Copenhagen, Jagtvej 128, 2200 Copenhagen, Denmark}
%\affiliation{Indian Institute of Astrophysics, 100 Feet Rd, Santhosapuram, Koramangala, Bengaluru 560034, India}
%
 
%===============================================
\begin{abstract}
We present a procedure  for calculating the  heating  of, and the infrared emission from, dust in a homogeneous spherical shell surrounded by a spherically symmetric source of radiation. The results are applicable to newly formed dust  either in supernova ejecta or in the circumstellar medium that has been swept up by the expanding shock wave. They can also be applied to the heating and IR emission from  dust in clumps or clouds  embedded in a homogeneous radiation field.  
 \end{abstract}

%===============================================
%==============================
\section{Introduction}
\label{introduction}
%==============================
A recurring astrophysical scenario is that of a  spherically symmetric dusty shell that is heated by an external spherically symmetric source of radiation. Examples are  dust in the expanding ejecta of a supernova (SN) or  dust in the surrounding circumstellar medium (CSM) that is  heated by the  radiation from an expanding shock wave, or a spherical dusty clump embedded in a homogeneous radiation field \citep[e.g.][]{sarangi18,sarangi22}. 
Here we present a procedure for calculating  the  heating of dust behind a spherically symmetric shock. The analytical equations presented here provide useful insight into the many variables of the problem, and in particular the morphology of the shell for different size cavities and shock proximity. We currently ignore the scattering of the incident radiation in the shell. 

Figure~1a  depicts a spherical dusty shell of radius $R_2$, containing a spherical cavity of radius $R_1$,  surrounded by a radiating shock at distance $R_{\rm s}$ from the center, $O$, of the shell. We consider the intensity of radiation seen by a dust grain located at position $A$ at distance $r$ from $O$. For evaluating the radiation incident at point $A$, we adopt a spherical coordinate system centered at $A$, with the polar axis oriented vertically in the $P$ direction as shown in the figure.

%==============================
\section{The intensity of the radiative shock}
\label{math}
%==============================

We first calculate the unattenuated flux incident on the grain  arriving in a cone at angle $\theta$ from the polar axis.  We assume that the nebula is heated by a shock with a  specific luminosity $L_{\nu}(\lambda)$, at wavelength $\lambda$, where
 %------------------- Lnu
\begin{eqnarray}
\label{lnu}
L_{\nu}(\lambda) & = &  4 \pi R_{\rm s}^2 \, \ell\, \epsilon_{\nu} (\lambda,T) \qquad \qquad  \qquad \  {\rm for \ an\  optically\  thin\  shock}  \nonumber \\
 & = &  4 \pi R_{\rm s}^2  \pi\, B_{\nu}(\lambda,T) \qquad  \qquad \qquad {\rm for \ an\  optically\  thick\  shock} \ ,
\end{eqnarray}
where $\ell = \overline {BD}$ and  $\epsilon_{\nu}(\lambda,T)$ [erg~s$^{-1}$~cm$^{-3}$~Hz$^{-1}$] are, respectively, the radial thickness and specific emissivity of the shock in the optically thin case. The specific luminosity and related quantities are  temperature dependent. 
 The specific intensity, or brightness of the shell $I_{\nu}(\lambda)$ [erg~s$^{-1}$~cm$^{-2}$~ster$^{-1}$~Hz$^{-1}$], is given by 
 \begin{equation}
\label{Inu}
I_{\nu}(\lambda) = \left({1 \over 4 \pi}\right)\, {L_{\nu}(\lambda) \over 4 \pi R_{\rm s}^2}
\end{equation}

The unattenuated specific flux incident on the grain, $F_{\nu}$[erg~s$^{-1}$~cm$^{-2}$~Hz$^{-1}$] , is
%------------------- Fnu0
\begin{eqnarray}
\label{fnu0}
F_{\nu}(\lambda) &  = &  \Omega\, \left[I_{\nu}(\lambda) / \cos \alpha\right]\ \  \qquad  {\rm for \ an\  optically\  thin\  shock} \qquad \nonumber \\
 & =  & \Omega\, I_{\nu} (\lambda)  \ \qquad \qquad  \qquad {\rm for \ an\  optically\  thick\ shock}  \ ,
\end{eqnarray}
where $\Omega=\sin \theta\, d\theta\, d\phi$  is a unit solid angle of the cone. In the optically thin case the  $\cos \alpha$ factor takes into account that the  path length $\ell'$ traversed through the shock is always larger than its radial thickness, and is given by $\ell' \approx \ell/\cos \alpha $, where $\alpha= \angle ABO$. 
The angle $\alpha$ is a function of $r$, $\theta$, and $R_{\rm s}$ and given by 
%------------------- cosA
\begin{equation}
\label{cosA}
\cos \alpha = [1- (r \sin \theta/R_{\rm s})^2]^{1/2}  \ .
\end{equation}

In the optically thick case, the incoming flux does not depend on the angle $\alpha$, since  the $\cos \alpha$ reduction in the flux projected into the cone is offset by the factor of $1/\cos \alpha$ increase in the shock area seen by the dust.  

The unattenuated specific flux incident on the grain is given by 
%------------------- Fnu0
\begin{eqnarray}
\label{fnu0}
F_{\nu}(R_{\rm s},r,\theta, \phi, \lambda,\tau=0) & =&   {\Omega\over 4 \pi}\, {\epsilon_{\nu} ( \lambda)\ell \over [1- (r \sin \theta/R_{\rm s})^2]^{1/2}} \ \  \qquad \qquad  {\rm for \ an\  optically\  thin\  shock}  \ , \nonumber \\
 & = &  {\Omega\over 4 \pi} {L_{\nu} (\lambda)\over 4 \pi R_{\rm s}^2} \,  \  \ \ \qquad \qquad \qquad  \qquad \qquad {\rm for \ an\  optically\  thick\  shock}  \ .
\end{eqnarray}

%==============================
\section{Extinction in the dusty shell}
\label{ext}
%==============================

The attenuated  flux incident on the dust from the  $\theta$ direction is 
%------------------- Fnu
\begin{equation}
\label{Fnu}
F_{\nu}(R_{\rm s},R_1,R_2,r,\theta,\phi,\lambda,\tau) =  F_{\nu}(R_{\rm s},r,\theta,\phi,\lambda,\tau=0)\ e^{-\tau(R_1,R_2,r,\theta, \phi, a, \lambda)} \ ,
\end{equation}
The optical depth,  $\tau$,  along the path, $t$, is given by
%------------------- tau
\begin{equation}
\label{tau}
\tau(R_1,R_2,r,\theta,\phi,a,\lambda) = \kappa(a,\lambda)\, \rho_{\rm d} \ t(R_1,R_2,r,\theta, \phi) \ ,
\end{equation}
where $\kappa(a,\lambda) = \sigma_{\rm gr}(a,\lambda)/m_{\rm gr}$  is the mass absorption coefficient of the dust at wavelength $\lambda$, $\sigma(a, \lambda) = \pi\, a^2\, Q_{\rm abs}(a,\lambda)$ is the grain's cross section, $Q_{\rm abs}$ is the absorption coefficient of the dust,   $m_{\rm gr}=4\pi\rho_{\rm gr} a^3/3$ is the grain's mass, $\rho_{\rm gr}$ is its mass density, $a$ its radius, and  $\rho_{\rm d}$ is the mass density of the dust in the shell.

The path length  $t\equiv \overline {AE}$  (Figure~1b) may, or may not, traverse the cavity, depending on the direction of  the dust viewing angle $\theta$.
The critical angle $\theta_{\rm c}$, which defines the boundary between the two cases, is
\begin{equation}
\label{thetac}
\theta_{\rm c}=\pi - \arcsin (R_1/r) \qquad .
\end{equation}
 
The path length through the nebula is then given by 
%------------------- tt
\begin{eqnarray}
\label{tt}
t (R_1,R_2,r,\theta,\phi)& = & -r \cos \theta + \left(R_2^2-h^2\right)^{1/2}  \qquad \ \ {\rm for}\  \theta < \theta_{\rm c}  \nonumber \\
 & = &  -r \cos \theta + \left(R_2^2-h^2\right)^{1/2}    \ \ 	  \qquad {\rm for}\  \theta > \theta_{\rm c}  \\
  &  & - 2 \left( R_1^2- h^2 \,  \right)^{1/2} \nonumber \ .
\end{eqnarray}
where  $h=r\, \sin \theta$.

Figure~1c shows the path length in units of $R_2$  as a function of angle $\theta$. The thin green lines depict the path length in the absence of a cavity.  At  $r=0$, the grain is at the center of the shell, and with no cavity all path length are independent of $\theta$ and equal to $R_2$.  The presence of a cavity, taken to have a radius $R_1=0.90\, R_2$, leaves the path length at all angles below $\theta_{\rm c}$ unchanged. However, as shown by the black lines, path lengths  are shortened when $\theta \ge \theta_{\rm c}$. The cusps in the figure correspond to the values of $\theta_{\rm c}$ for the different values of $r$.  The red and blue lines correspond  to the path lengths when $r=R_1$ and $R_2$, respectively.

%==============================
\section{Temperature of the dust}
\label{temp}
%==============================

The total power, $P(R_1, R_2, r,\tau)$,  absorbed by the grain is given by integrating the incident attenuated specific flux  from all directions and  frequencies  times the grain's cross section
%------------------- pow
\begin{equation}
\label{pow}
P(R_1,R_2,r,\tau)= 2 \pi\, \int_0^{\infty}  \sigma_{\rm gr}(a,\lambda)\  \int_0^{\pi} \ F_{\nu}(R_{\rm s},R_1,R_2,r,\theta,\lambda,\tau)\, \sin \theta\, d\theta \, d \nu \ .
\end{equation}

The  temperature of a dust grain , $T_{\rm gr}$, is obtained by equating its heating rate, $P$, to its luminosity, $L_{\rm gr}$,
%------------------- temp
\begin{equation}
\label{temp}
 L_{\rm gr} = 4 m_{\rm gr}\ \int_0^{\infty} \ \pi B_{\nu}\left(\lambda, T_{\rm gr}(r)\right)\, \kappa(a,\lambda)\, d\nu \ ,
\end{equation}
where $B_{\nu}$ is the Planck function in erg~s$^{-1}$~cm$^{-2}$~sr$^{-1}$~Hz$^{-1}$.

Figure~1d depicts the radial profile of the dust temperature with a cavity (dashed lines) and without a cavity (solid lines). Calculations were performed for 0.1 $\mu$m radius silicate grains, heated by an optically  thin shock at a distance $R_{\rm s} = 5 R_2$, with a luminosity of $10^8$~L$_{\odot}$ and thickness $\ell = 0.1 R_{\rm s}$, where $R_2 = 1\times 10^{16}$~cm. The temperature profiles are plotted for different radial optical depths at $\lambda (V)= 0.55\, \mu$m. The radial optical depth at any wavelength $\lambda$ is  given by $\tau(\lambda)=\kappa(\lambda)\, \rho_{\rm d}\, R_2$. Larger optical depths require an increase in the density of dust grains in the shell. Densities were chosen to give  optical depths equal to 1.0, 3.0, 10.0, and 30.0. For $\tau(V)=1$ the resulting density and the dust mass of the filled shell were $4.0\times 10^{-20}$~g~cm$^{-3}$ and $8.5 \times 10^{-5}$~M${_\odot}$, respectively. The larger dust densities correspond to dust masses of  $2.5 \times 10^{-4}$,  $8.5 \times 10^{-4}$, and $2.5 \times 10^{-3}$~M${_\odot}$.
With no cavity, the dust temperature exhibits the obvious rise towards the surface, where the incident radiation is less attenuated. The same general trend is seen in the presence of a cavity. However the presence of a cavity has the effect of increasing the radiation reaching subshells with radii $> R_1$, so that the resulting temperature profile is somewhat higher and slightly rises towards the edge of the cavity.  This effect is more pronounced at lower optical depths.
The pink horizontal line marked $\tau(V)=0$ depicts the temperature of a grain located at  $R_2$, that is heated by the shock if it were a point source in the center of the shell. 

%==============================
\section{Infrared Emission From the Shell}
\label{emm}
%==============================

The specific infrared (IR) emissivity in the shell is 
\begin{equation}
\label{epsIR}
\epsilon_{\nu}(r,a,\lambda,T_{\rm gr}(r)) = 4\, \rho_d\,   \pi B_{\nu}(\lambda, T_{\rm gr}(r))\, \kappa(a,\lambda) \ ,
\end{equation}
and the specific infrared (IR) luminosity of the shell is given by the integral  
%------------------- Lnu
\begin{equation}
\label{Lnu}
L_{\nu}(a,\lambda) = 4 \pi \ \int_{R_1}^{R_2}\  \epsilon_{\nu}(r,a,\lambda,T_{\rm gr}(r)) \ r^2  \, dr \ .
\end{equation}

Figures~1f and 1g show the internal (intrinsic) specific luminosity for selected subshells $\{r,r+dr\}$ as a function of wavelength for visual optical depths of $\tau(V)$=1 (Fig.~1f), and $\tau(V)$=10 (Fig.~1g). The blue and red curves  correspond to the specific luminosities from the subshells without, and with a cavity, respectively. The thick blue line and the thick dashed red line  represent the total specific luminosity, summed over all subshells. The cavity in the two figures was characterized by a radius $R_1 = 0.90\, R_2$.  

The internal specific luminosity in the subshells is characterized by a colder spectrum and lower intensity with increasing optical depth. However, the total spectrum is dominated by the emission from the outer subshells. As a result, at high optical depths, the total spectrum is not greatly affected by the presence of a cavity (Fig. 1g), since the incoming radiation is predominantly absorbed in the outer subshells. At lower optical depths, (Fig. 1f) colder subshells contribute to the total internal spectrum for the cavity free case. The resulting total spectrum is therefore colder compared to that of the shell with a cavity where only the hot dust in the outer subshells contributes to the internal energy.

%The figure also shows that the total internal energy  increases with optical depth since a larger fraction of the incoming radiation is absorbed in the shell.

%==============================
\section{The Escape of the Thermal Emission From the Shell}
\label{esc}
%==============================

The observed infrared emission from the dust is the fraction of the total internal energy density that escapes the shell. 
The escape probability of photons from a spherical homogeneous shell was recently presented by \cite{dwek24}. In that study, the dust emissivity in the shell was assumed to be constant, whereas in the present case it has a radial dependence. The procedure for calculating the escaping radiation  is essentially the reverse of that used to calculate the flux of incident radiation that heats the dust. Let $\epsilon_{\nu}(r,a,\lambda,T_{\rm gr}(r))$ be the specific energy density at point $A$, at distance $r$ from the center of the nebula. The specific luminosity escaping from the shell is given by integrating the escaping emission in all directions, and over all subshells
\begin{equation}
\label{ }
L_{\nu}(R_1, R_2,a,\lambda)  =  \int_{R_1}^{R_2}\  r^2\, \epsilon_{\nu}(r,a,\lambda,T_{\rm gr}(r))\ \left[ 2 \pi\,  \int_0^{\pi} \  e^{-\tau(R_1,R_2,r,\theta,a,\lambda) } \sin \theta d\theta\right] dr \ ,
\end{equation}
which for $\tau = 0$ reduces to equation (\ref{Lnu}).

Figures~1i and 1j represent the total internal and escaping specific luminosity for different optical depths. The solid lines are the internal flux in the shell, and the dotted lines represent the escaping flux.  The figures  show that the total internal luminosity increases with optical depth, since more of the incident radiation is absorbed by the shell.  The figure also shows that the silicate  features in the escaping spectrum are suppressed but still detectable at low optical depths, $\tau(V) < 10$, and significantly weaker at larger ones.  Suppression of these  features to create a smooth emerging spectrum requires any emission to be self-absorbed at these IR wavelengths. This requires the dust temperature to be isothermal throughout the subshell that contribute significantly to the emission, so that any absorption or emission features emerging from  a subshell are cancelled out by the corresponding emission or absorption in neighboring subshells. The strength of the silicate features will depend on the temperature gradient in the shell and the subshells that are the dominant contributors to the escaping flux. 

Figures 1e, 1h, and 1k are the equivalent of Figs. 1d, 1g, and 1j, but for a shell that is heated by a shock located at a short distance of $R_s = 1.2 R_2 = 1.2\times 10^{16}$~cm from the center of the shell. Other shock parameters, its total luminosity and  thickness, are identical to the previous case. The figures show that because of the shock's proximity a larger fraction of its luminosity is actually absorbed by the shell. Because the shock's luminosity emerges from a smaller area, its brightness, $I_{\nu}$, is larger, resulting in an increase in dust temperature in all subshells, an effect that is manifested in their spectra, and in that of the escaping radiation. 

%==============================
\section{Summary}
\label{sum}
%==============================

We presented the basic mathematical outline for calculating the internal spectrum from a spherically symmetric shell that is heated by an external spherically symmetric source of radiation. The scenario envisioned here is that of an expanding radiative shock that is heating an interior shell of either ejecta or cooled, swept up, CSM dust. The presence of a cavity affects the emerging radiation, depending on its size. The results are  significantly affected by the shock's proximity to the shell, or its luminosity. 
Our study highlights the interplay between the various global parameters characterizing the shock and the shell. It also shows the many potential degeneracies in fitting any observed spectrum, in which changes in the shock  luminosity or distance, in the total dust mass or cavity size, as well as changes in dust composition and size distribution can yield similar results.   
The results presented here neglect the effect of scattering of the incoming radiation by the dust. Scattering will have an effect on the fraction of the incident UV-optical radiation that penetrates the shell, and on its trajectory and penetration depth. It will have little effect on the escape of the infrared emission from the heated dust. We also neglect the additional  heating of the dust by the outgoing IR radiation. Any contribution to the observed emission from collisionally heated dust embedded in the hot dense post-shock gas is assumed to be negligible.

%===============================
% Figure 1 
%===============================
\begin{figure*}[p]                                                                               %    <====== positioning of figure
\centering
\includegraphics[width=5.0in]{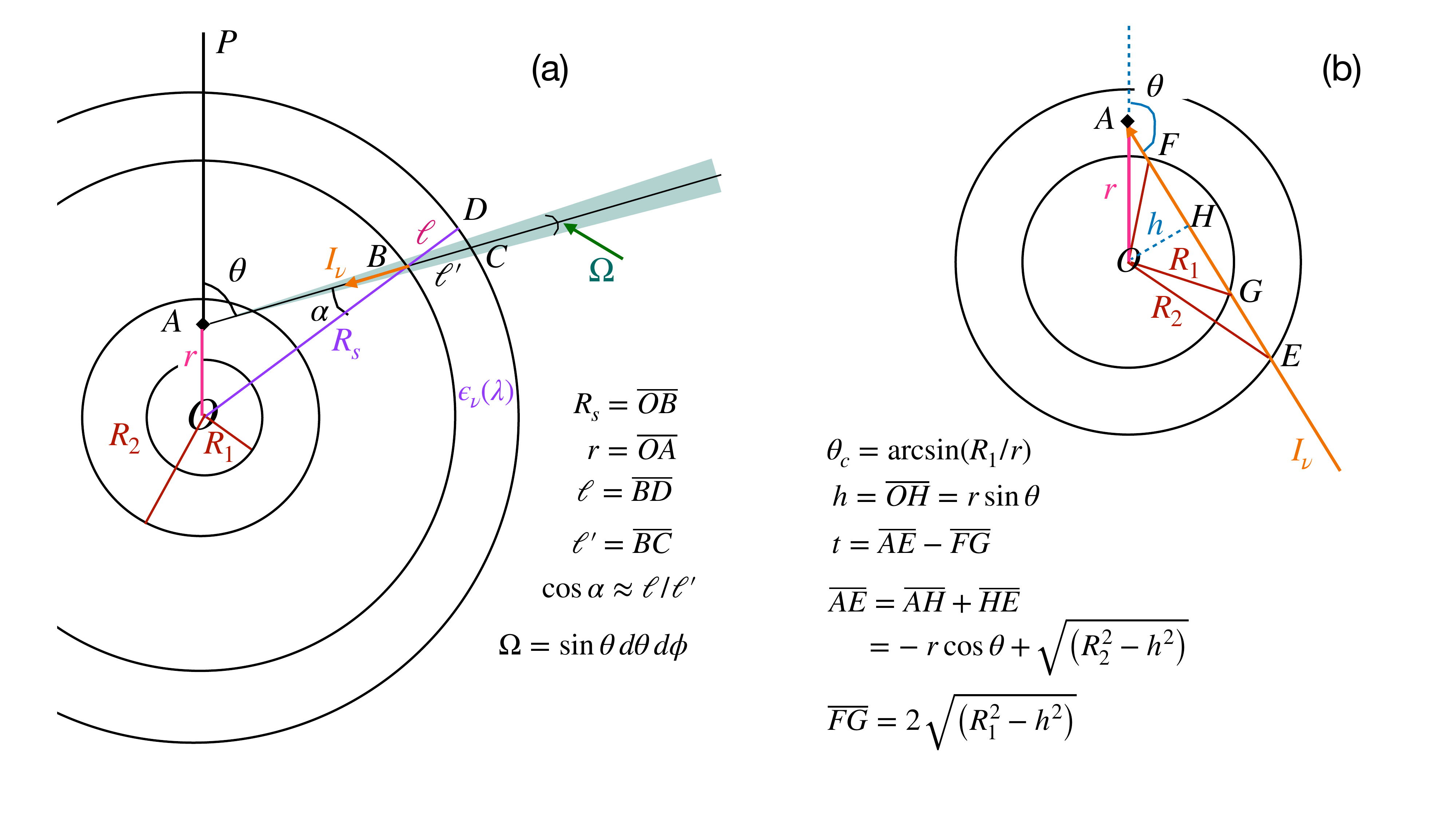} \\
~~\\
\includegraphics[width=2.0in]{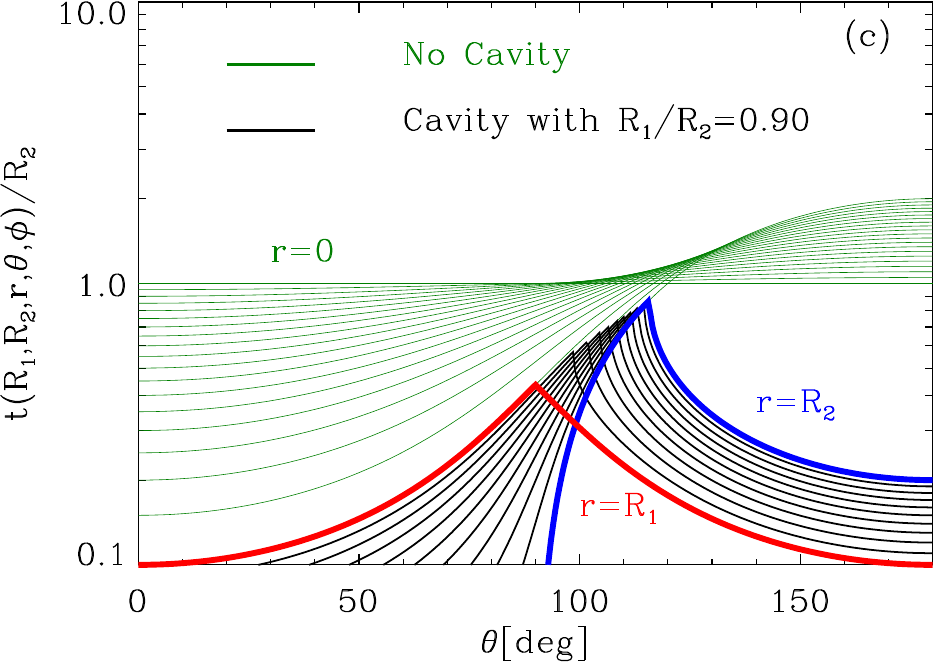} 
\includegraphics[width=2.0in]{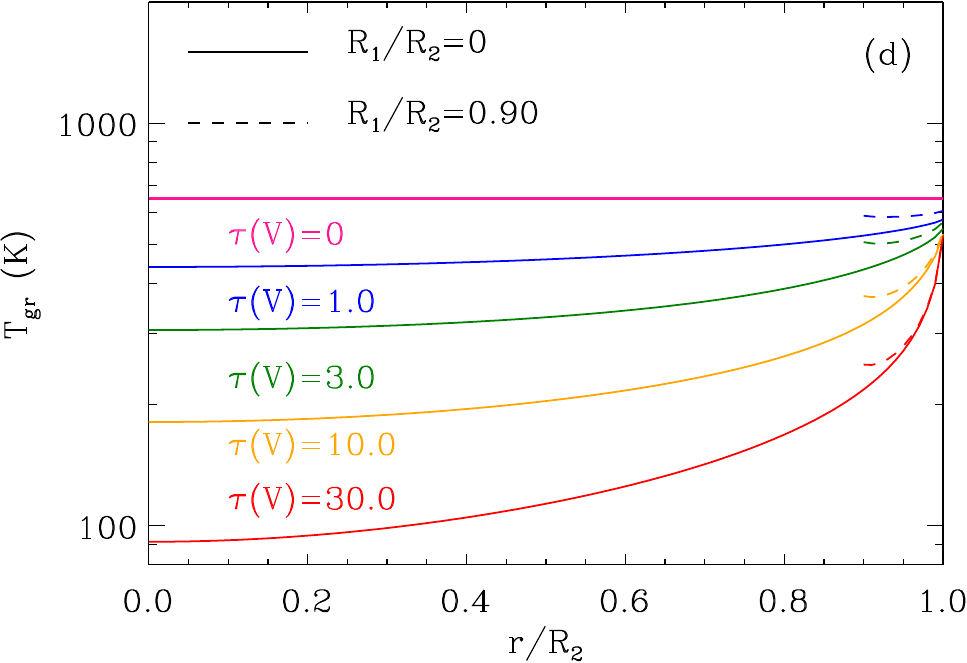} 
\includegraphics[width=2.0in]{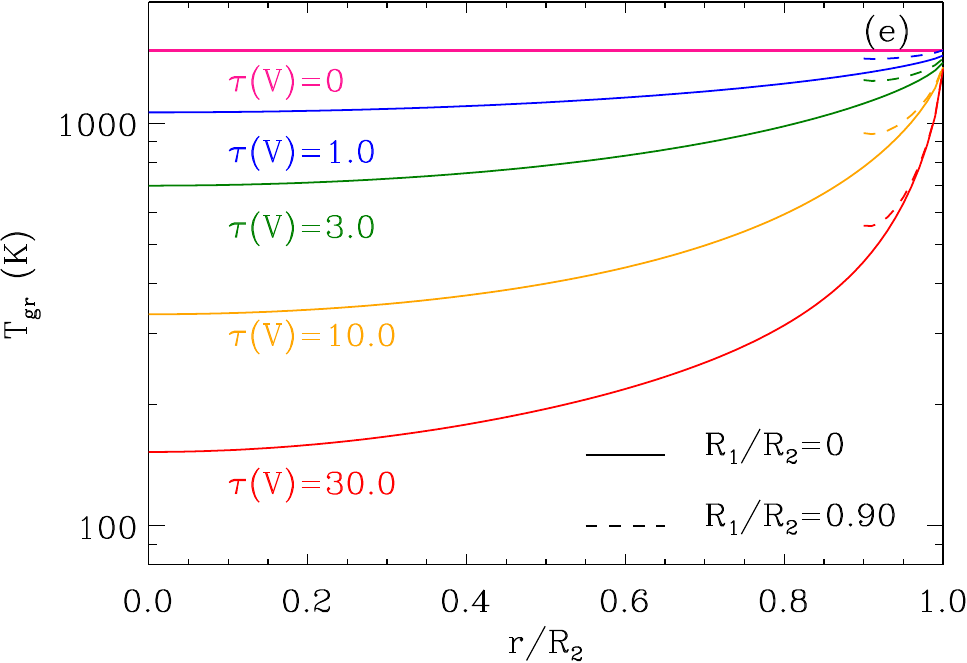}\\
~~\\
\includegraphics[width=2.0in]{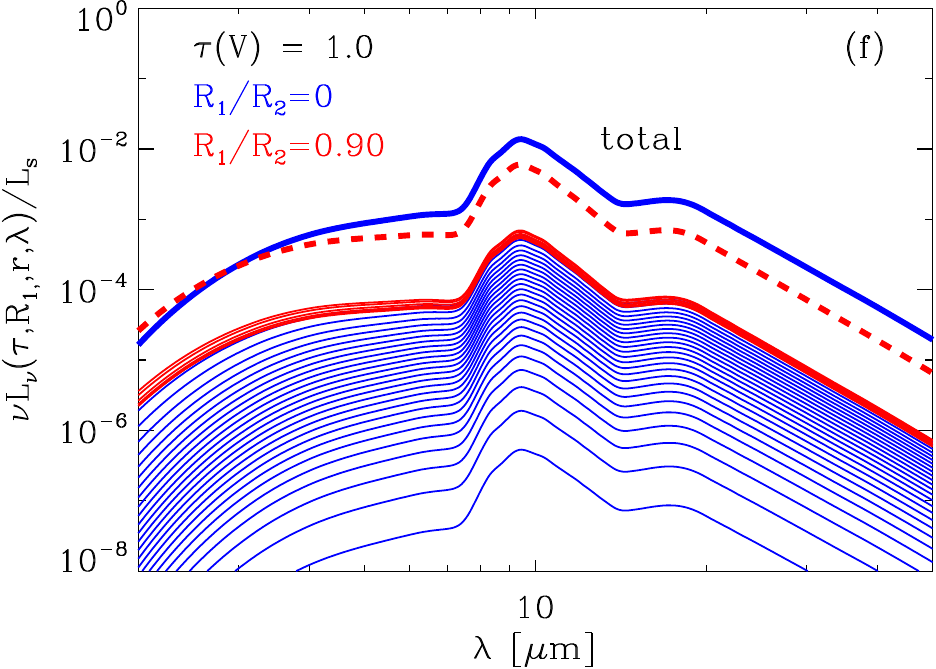} 
\includegraphics[width=2.0in]{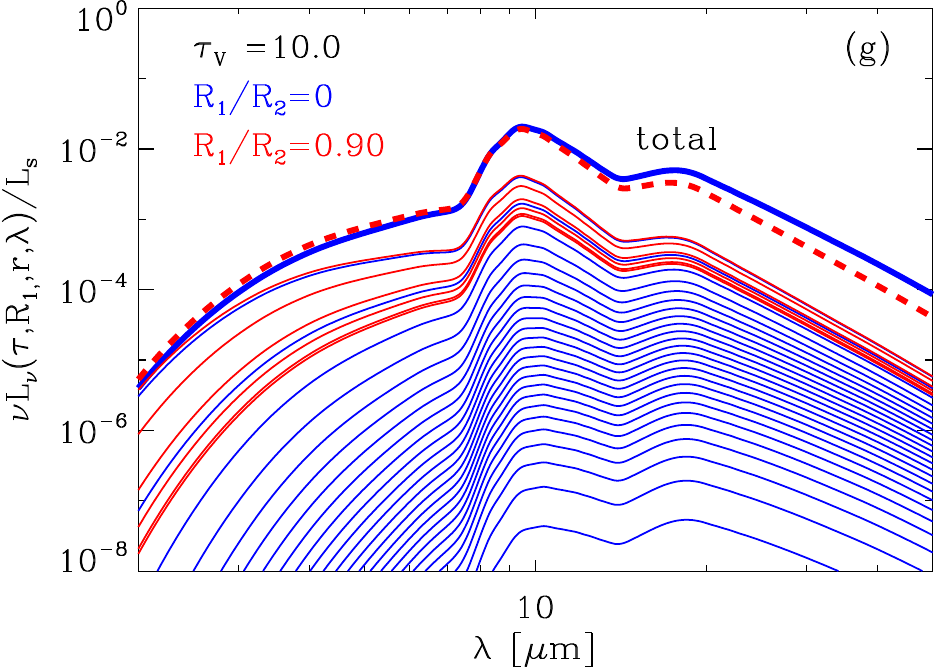} 
\includegraphics[width=2.0in]{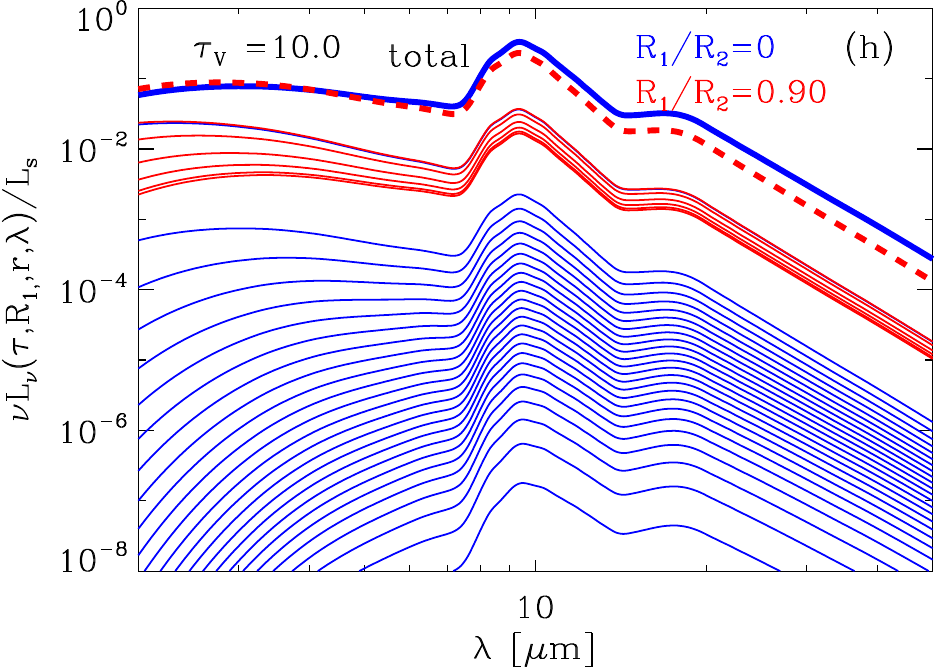} \\
~~\\
\includegraphics[width=2.0in]{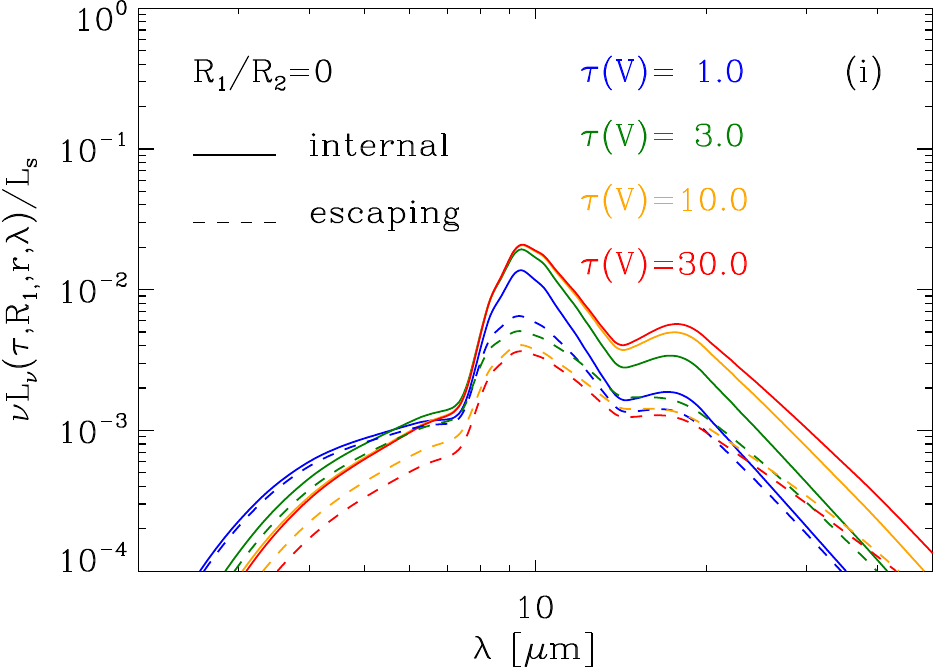} 
\includegraphics[width=2.0in]{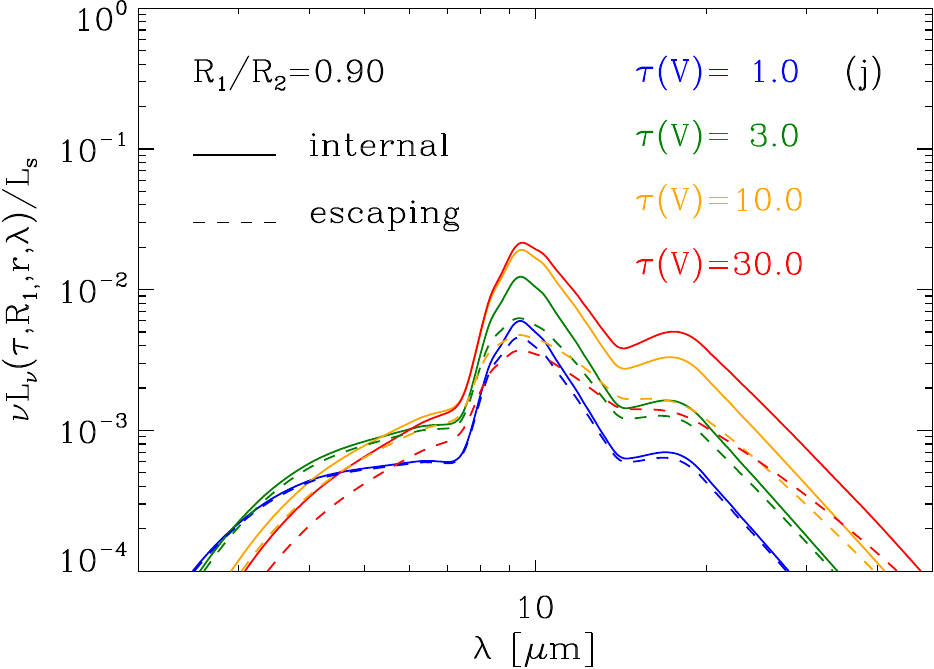} 
\includegraphics[width=2.0in]{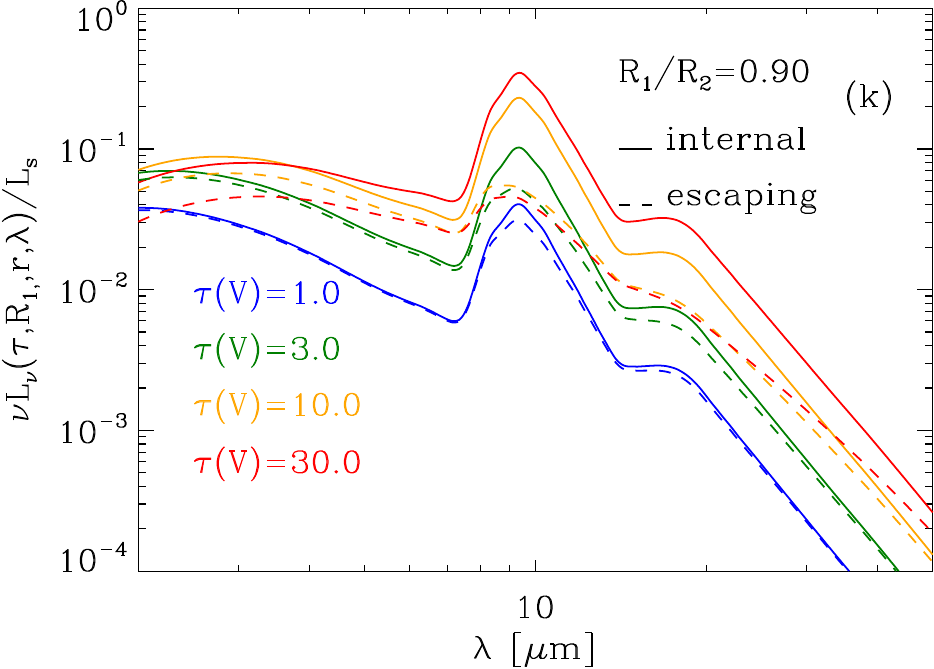} 
\caption{ (a)-(b) Geometrical presentation; (c) Path length through shell; (d, f, g, i, j) model results for shock distance of $R_{\rm s}=5.0\times 10^{16}$~cm; (e, h, k) model results for $R_{\rm s}=1.2\times 10^{16}$~cm. For sake of clarity, fewer subshells were plotted for the cavity case in (c, f, g, h). Details in text.}
\end{figure*}

%---------------------------------~
%\input{Table_1.tex}
%----------------------------------

\begin{acknowledgments}
E.D. thanks the Niels Bohr Institute for their hospitality during the completion of the manuscript, which was supported by a VILLUM FONDEN Young Investigator Grant (project number 25501).  Work by R.G.A. was supported by NASA under award number 80GSFC21M0002.
\end{acknowledgments}

\bibliography{/Users/elidwek/SCIENCE/003-Bib_Desk/Astro_BIB.bib}

 \end{document}